\documentclass[prl,twocolumn,tightenlines,showpacs,byrevtex,11pt]{revtex4}

\usepackage{graphicx}
\usepackage{dcolumn}
\usepackage{bm}

\begin{document}
\preprint{ver 7}
\title{Clock Synchronization based on Second-Order Quantum Coherence of Entangled Photons}
\author{Thomas B. Bahder}
\email{bahder@arl.army.mil}
\author{William M. Golding}
\affiliation{
Army Research Laboratory \\
2800 Powder Mill Road \\
Adelphi, Maryland, USA 20783-1197}
\date{\today}

\begin{abstract}
We present an algorithm for synchronizing two clocks based on second-order
quantum interference between entangled photons generated by parametric
down-conversion. The procedure is distinct from the standard Einstein two-way
clock synchronization method in that photon correlations are used to define
simultaneous events in the frame of reference of a Hong-Ou-Mandel (HOM)
interferometer. \ Once the HOM\ interferometer is balanced, by use of an
adjustable optical delay in one arm, arrival times of simultaneously generated
photons are recorded by each clock.\ Classical information on the arrival
times is sent from one clock to the other, and a correlation of arrival times
is done to determine the clock offset.

\end{abstract}

\maketitle

High-accuracy synchronization of clocks plays an important role in fundamental
physics and in a wide range of applications such as communications, message
encryption, navigation, geolocation and homeland security. \ There are two
classical methods of time synchronization of spatially separated clocks:
\ Eddington slow clock transport\cite{EddingtonClockTransport}, and Einstein
synchronization\cite{LLClassicalFields} by exchange of light signals. \ \ In
Eddington slow clock transport, two co-located clocks are initially
synchronized, and then one of these clocks is slowly transported \ to another
location to synchronize a distant clock. \ For most technological applications
today, this method is not practical because it requires transport of hardware
as well as conflicting requirements: on the one hand clock transport must be
slow to reduce the effect of time dilation, but on the other hand it must be
fast enough so that significant time differences do not accrue from
unavoidable timing errors due to the limited frequency stability of the
transported clock's mechanism or due to gravitational potential differences
along the path of the transported clock\cite{LLClassicalFields}. \

Einstein synchronization consists of a round-trip exchange of classical light
signals between two spatially separated clocks\cite{LLClassicalFields}. A
signal is sent from clock B (which is to be synchronized with clock A) and
reflects off the face of clock A, and then returns to clock B. \ The signal
that reflected off the face of clock A carries classical information, from
clock A to clock B, consisting of the time on clock A at the event of signal
reflection. In practical applications today, a\ satellite system, such as the
Global Positioning System (GPS), is used for synchronizing two spatially
separated clocks\cite{ParkinsonSpilker1996,Kaplan96,Hofmann-Wellenhof93}.
\ The GPS is a satellite system based on Einstein synchronization, since
signals are sent from satellite-to-ground and from ground-to-satellite to
synchronize the satellite clocks with a master clock on Earth\cite{Bahder2003}%
. The time-synchronization accuracy provided by a GPS receiver is on the order
of 20 ns. \ \ However, there are many applications, such as coherent detection
of electromagnetic signals, where time synchronization is required to an
accuracy that cannot be provided by the GPS. \ Recently, there have been
several clock synchronization schemes proposed that are based on quantum
mechanical
principles\cite{Chuang2000,Jozsa2000,Giovannetti2001,Giovannetti2002,ShihPRL2003}%
. The hope is that quantum mechanical methods can provide a higher clock
synchronization accuracy than classical methods.

In this letter, we present a simple quantum mechanical algorithm to
synchronize two spatially separated clocks to an accuracy limited by the
resolution of the Hong-Ou-Mandel (HOM) \ interferometer, which has been
demonstrated to be better than one hundred femtoseconds\cite{HOM}. Our
algorithm for clock synchronization has four key features. First, no
information is needed on the geometric distances between clocks. Second, the
algorithm works in vacuum because it does not rely on the presence of an
optical medium, and hence is applicable to space-based systems.  Third,
rather than using massive particles\cite{Jozsa2000}, the algorithm is based on
entangled photon pairs, which are weakly coupled to the environment so the
synchronization can be carried out over large distances. Finally, the
essential elements that are contained in the algorithm, such as single-photon
detection, time of arrival tagging, and signal time delay techniques, have
already been demonstrated in the laboratory at various levels of precision.

\begin{figure}
\includegraphics{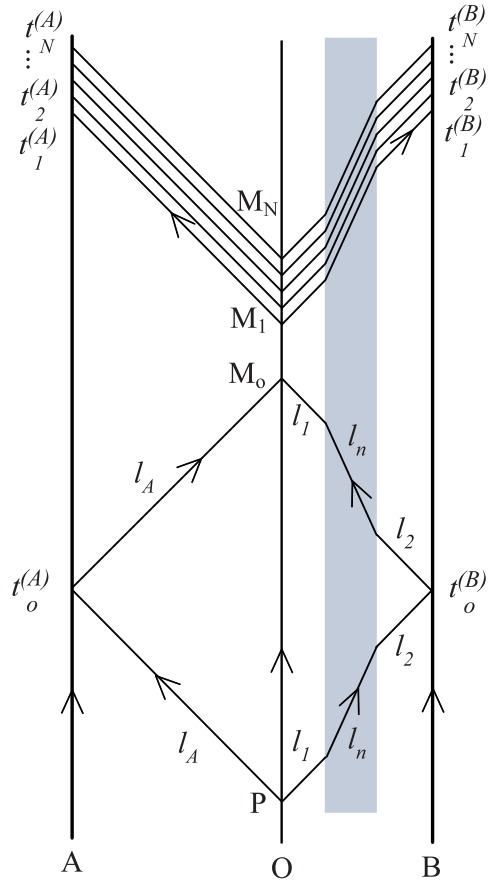}
\caption{\label{ClockSynchSpaceTimev3}
The space-time diagram shows the world lines of clocks A and B, the
world line of the entangled photon distribution center, \textit{O}, and the
co-located HOM interferometer. The interferometer is balanced at event M$_{o}%
$. Entangled photon pairs are launched at events M$_{1\text{,}}$...,M$_{N}$.
Arrival proper times, $\tau_{i}^{(A)}$ and $\tau_{i}^{(B)}$ , $i=1\cdots N$,
are recorded at clock A and B, respectively.}
\label{clockAcorrection1}
\end{figure}

Our synchronization algorithm uses photon coincidence counting in the frame of
reference in which the entangled photon source is at rest\cite{InertialFrame}.
The underlying method is based on second order quantum interference between
two entangled photons, rather than on classical exchange of information, as
used in classical Einstein synchronization. The synchronization algorithm\ is
shown schematically in the Minkowski diagram in Figure~\ref{ClockSynchSpaceTimev3}.
The goal is to
synchronize clock B with clock A. \ Clocks A and B are assumed stationary in
the frame of reference in which the source of correlated photons is at rest.
The HOM\ interferometer is spatially co-located with the photon source, and
both are on world line \textit{O.} \ Also shown in
Figure~\ref{ClockSynchSpaceTimev3} in grey is the
world volume of an optical medium which has an adjustable index of refraction,
$n$, which is located between the world line \textit{O} and world line of
clock \textit{B}.

We use \textquotedblleft hardware time\textquotedblright, $\tau^{\ast}$,\ \ to
mean the elapsed time (since some epoch) kept by a real hardware
clock\cite{ClockDefinition}. Proper time, $\tau$, is the elapsed time that is
kept by an ideal clock and it depends on the world line, or history, of the
ideal clock. The difference between \textquotedblleft hardware
time\textquotedblright\ and proper time is that a real hardware clock can have
mechanical imperfections that make \textquotedblleft hardware
time\textquotedblright\ deviate from proper time. \ On the other hand,
coordinate time, $t$, is a global quantity, which is associated with the
metric of space-time, $g_{ij}$, and enters in the definition of the system of
4-dimensional coordinates. \ It is coordinate time that provides the
connection between proper times on two different world lines. However, it is
well-known that coordinate time is not a measureable
quantity\cite{Pirani57,Synge1960,Soffel89,Brumberg91,Guinot97}. \

A synchronization algorithm must deal with synchronization of time kept by
real hardware clocks, which means it must deal with \textquotedblleft hardware
time\textquotedblright. In what follows, we assume that our hardware clocks
are sufficiently stable so that \textquotedblleft hardware
time\textquotedblright\ \ is a good approximation of proper time: $\tau^{\ast
}=\tau$, over the times of interest in the synchronization procedure. \ We
also assume that both clocks A and B have the same rate with respect to
coordinate time, and with respect to each other\cite{CoordinateTimeChoice}%

\begin{equation}
\frac{d\tau^{(A)}}{dt}=1=\frac{d\tau^{(B)}}{dt} \label{TimeRates1}%
\end{equation}
so we are neglecting the effect of gravitational potential differences between
the locations of clock A and B during the synchronization time. Therefore, we
assume that both clocks A and B keep proper time, but they have an unknown
constant offset between them, which we seek to determine by our
synchronization algorithm. \

The algorithm is as follows. Correlated photon pairs are generated
in a crystal by parametric
down-conversion~\cite{Burnham1970,Ghosh1986,KlyshkoBook1988,Rubin1994}
at event\ \textit{P}. \ The pairs are created simultaneously at a
unique event in space-time, within the crystal to an accuracy of
better than 100 femtoseconds\cite{HOM}. \ \ One photon from the
pair arrives at clock A at coordinate time\ $t_{o}^{(A)}$, is
reflected back toward the world line \textit{O}, and arrives at
event $M_{o}$. \ The other member of the photon pair travels
through an adjustable optical medium to clock B, shown in grey in
Figure 1, and is reflected back toward world line \textit{O},
again traveling through the optical medium. \ This procedure is
continued while the index of refraction $n$\ of the optical medium
is adjusted, until both photons from the pair arrive at a single
event $M_{o}$. The event $M_{o}$ is identified by a minimum
observed in the HOM\ interferometer, indicating that the
interferometer is balanced. \ The balance condition means that the
\textit{coordinate time} of photons arriving at clock A and B are
equal,
\begin{equation}
t_{o}^{(A)}=t_{o}^{(B)} \label{ZeroCoordinateTime}%
\end{equation}
indicating that these events are simultaneous in the system of coordinates in
which the interferometer is at rest\cite{InertialFrame}. We define
simultaneous events as events with the same coordinate time in the given
system of inertial coordinates\cite{LLClassicalFields}. \ Once the
interferometer is balanced, correlated photon pairs are emitted on the world
line \textit{O} at events $M_{1}$, ... $M_{N}$. \ Photons detected at clock B
have passed through the optical device, whose delay has been adjusted to
obtain the balanced interferometer condition. \ Data consisting of the arrival
time of photons at clock A and B is recorded on the world lines of clocks A
and B, consisting of the sets of numbers, $\{\tau_{i}^{(A)}\}$ and $\{\tau
_{i}^{(B)}\}$,\ $i=1\cdots N$, as measured with respect to clocks A and B,
respectively. \ We use the notation $\tau_{i}^{(A)}$ and $\tau_{i}^{(B)}$\ for
the proper times\ of photon arrival at clock A and B, as distinct from
coordinate time of photon arrival, $t_{i}^{(A)}$ and $t_{i}^{(B)}$. \

Photons that are \textquotedblleft coincident at clock A and clock
B\textquotedblright\ are defined to be those that are simultaneous in the
inertial system of space-time coordinates in which the HOM\ interferometer is
at rest. \ Our definition of simultaneous photon arrival times on world lines
A and B means that the coordinate times of photons arriving on world line A
and B are the same%
\begin{equation}
t_{i}^{(A)}=t_{i}^{(B)} \label{SimultaneousTime1}%
\end{equation}
for photons $i=1,...,N$. On the world line of clock A, the proper time elapsed
between the events at coordinate time of reception $t_{i}^{(A)}$ and
\ $t_{1}^{(A)}$ is given by%
\begin{equation}
\tau_{i}^{(A)}+\Delta\tau^{(A)}=t_{i}^{(A)}-t_{1}^{(A)}
\label{clockAcorrection}%
\end{equation}
\ where $\Delta\tau^{(A)}$\ is the clock correction that relates coordinate
time to proper time of clock A. \ A similar relation exists for clock B:%

\begin{equation}
\tau_{i}^{(B)}+\Delta\tau^{(B)}=t_{i}^{(B)}-t_{1}^{(B)}
\label{clockBcorrection}%
\end{equation}
where $\Delta\tau^{(B)}$\ is the clock correction that relates coordinate time
to proper time for clock B. \ It is clear from Eq.(\ref{SimultaneousTime1}%
)-(\ref{clockBcorrection}) that the proper times of clocks A and B are related
by
\begin{equation}
\tau_{i}^{(B)}-\tau_{i}^{(A)}=\Delta\tau^{(A)}-\Delta\tau^{(B)}
\label{ProperTimeRelations}%
\end{equation}
for all received photon pairs $i=1,...,N$.

\begin{figure}
\includegraphics{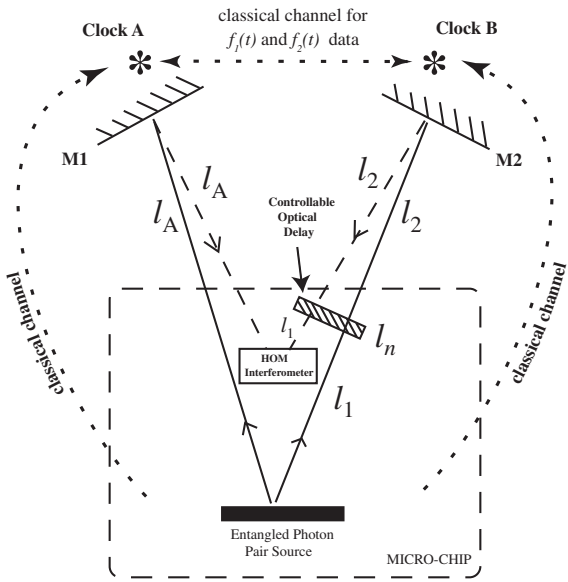}
\caption{\label{QuantClockSynch11}
The 3-dimensional laboratory view of the synchronization
method.  The equal time of propagation from the entangled photon
source to clocks A and B means that $l_{n}=l_{A}-(l_{1}+l_{2})$,
where $l_{n}$ is the optical path length of the delay line.}
\end{figure}

The data taken by clock A and clock B is a series of $N$ photon arrival times,
tagged at the detector, with respect to each clock's respective hardware
times, $\tau_{i}^{(A)}$ and $\tau_{i}^{(B)}$, $i=1,...,N$. Both clocks record
photon arrival time data for a time $T$ measured with respect to their local
clock. \ The time $T$ is long compared to the expected clock difference
between clock A and clock B. \ The photon arrival time data, $\{\tau_{i}%
^{(A)}\}$ and $\{\tau_{i}^{(B)}\}$, are assembled into functions $f_{A}(t)$
and $f_{B}(t)$:

\begin{equation}
f_{A}(t)=\frac{1}{\sqrt{N}}\sum_{i=1}^{N}\delta(t-\tau_{i}^{(A)})
\label{dataFunctionA}%
\end{equation}
where $\delta(t)$ is the Dirac delta function, and the analogous definition
for $f_{B}(t)$. \ Classical information consisting of the function $f_{A}(t)$
is transmitted from clock A to clock B via a classical information channel,
see Figure~\ref{QuantClockSynch11}. \ A correlation is then performed at clock B of the data
functions $f_{A}(t)$ and $f_{B}(t)$:%
\begin{equation}
g(\tau)=\int_{-\infty}^{+\infty}f_{A}(t)f_{B}(t-\tau)dt \label{CorrFunction}%
\end{equation}
Substituting the values of the data functions $f_{A}(t)$ and $f_{B}(t)$ into
Eq.(\ref{CorrFunction}) and doing the integral gives
\begin{equation}
g(\tau)=\frac{1}{N}\sum_{i=1}^{N}\sum_{j=1}^{N}\delta(\tau-\tau_{j}^{(A)}%
+\tau_{i}^{(B)}) \label{GFunctionSum}%
\end{equation}
For most values of $\tau$, the correlation function $g(\tau)$\ is $O(1/N)$.
However, the diagonal terms in the sums, for $i=j$, contribute $N$ terms at
the same value of $\tau$\ , since the value of the terms $\tau_{i}^{(A)}%
-\tau_{i}^{(B)}$\ $=\Delta\tau^{(B)}-\Delta\tau^{(A)}$\ , as given by
Eq.(\ref{ProperTimeRelations}), and the correlation function is of order unity
at a value of $\tau=\tau_{o}=\Delta\tau^{(B)}-\Delta\tau^{(A)}$. \ According
to Eq.(\ref{ProperTimeRelations}), if $\tau_{o}$ is added to the value of the
current clock time $\tau^{(B)}$, then clock B will be synchronized to clock A.
\ Note that the clock time on clock B after synchronization is, in general,
not equal to coordinate time in the chosen system of coordinates.

In summary, we have defined a simple algorithm to synchronize two real clocks
that are spatially separated. \ The accuracy of the algorithm is based on
accurate control of an optical delay line and on the accuracy of second order
quantum interference exhibited by correlated photons in a HOM\ interferometer,
which is below 100 femtoseconds\cite{HOM}. \ Each experimental element in the
algorithm, such as the source of correlated photons, photon detection, arrival
time tagging, and the HOM interferometer, have been separately experimentally
demonstrated during the past several years. \ Consequently, this algorithm is
today a viable method of synchronizing two spatially separated clocks. \ This
algorithm may be useful in applications, such as communications, message
encryption, navigation, geolocation systems, to achieve synchronization
accuracy that is beyond the current reach of classical synchronization methods.

\bigskip {\it Acknowledgement} This work was sponsored by the Advanced Research Development Agency (ARDA) and the  National Reconnaissance Office (NRO).  One of the authors (W. M. G.) performed part of this work while at the  Naval Research Laboratory.   The authors wish to acknowledge helpful discussions and suggestions from Pete Hendrickson.


\begin{thebibliography}{99}
\bibitem {CoordinateTimeChoice}This is a convenient choice of metric for the
flat space-time. \ \ Alternatively, we could have chosen a metric with
coordinate time $t^{^{\prime}}=\alpha t$, then $d\tau/dt^{^{\prime}}=1/\alpha$.
\end{thebibliography}
\end{document}